\documentclass[twocolumn,showpacs,preprintnumbers,amsmath,amssymb]{revtex4}

\usepackage{graphicx}
\usepackage{dcolumn}
\usepackage{bm}

\begin{document}

\title{A new cooperation mechanism of kinesin motors when extracting membrane tube}

\author{Ziqing Wang$^{1}$}
\email{wzqphy@yahoo.cn}

\author{Ming Li$^{2}$}
\email{liming@gucas.ac.cn}

\affiliation{${}^{1}$Institute of Biophysics and College of Science, Northwest A$\&$F University, Yangling  712100, China}

\affiliation{${}^{2}$College of Physical Science, Graduate University of Chinese Academy of Sciences, Beijing 100190, China}

\date{\today}

\begin{abstract}
Membrane tubes are important elements for living cells to organize many functions. Experiments have found that membrane tube can be extracted from giant lipid
vesicles by a group of kinesin. How these motors cooperate in extracting the fluid-like membrane tube is still unclear. In this paper, we propose a new cooperation
mechanism called  two-track-dumbbell model, in which kinesin is regarded as a dumbbell with an end (tail domain) tightly bound onto the fluid-like membrane and the
other end (head domain) stepping on or unbinding from the microtubule. Taking account of the elasticity of kinesin molecule and the exclude volume effect of both
the head domain and the tail domain of kinesin, which are not considered in previous models, we simulate the growth process of the membrane tube pulled by kinesin
motors. Our results indicate that motors along a single microtubule protofilament can generate enough force to extract membrane tubes from vesicles, and the average
number of motors pulling the tube is about $8\sim 9$. These results are quite different from previous studies (Ref. \cite{camp.08}), and further experimental tests
are necessary to elucidate the cooperation mechanism.

\end{abstract}


\maketitle

\section{Introduction} \label{S.In}

Membrane tubes widely exist in eukaryotic cells, such as in the endoplasmic reticulum and the Golgi apparatus etc.\cite{lee.88}, and play an important role in
intracellular transportation as a long-range tubular transport intermediate which is essentially different from the conventional notion of small vesicle
 transportation\cite{sciaky.97}. Experiments have revealed that the membrane tube in living cells always co-occur with microtubules \cite{allan.94},
 and recent experiments \textit{in vitro} have shown that both the processive and the non-processive motors can extract membrane tubes from lipid
 giant unilamellar vesicles \cite{roux.02,koster.03,leduc.04,shaklee.08,shaklee.10}. Theories and experiments also showed, after the formation of a membrane tube
from a vesicle, the force needed to keep the growth of the membrane tube is about $f_0\simeq 27.5 \pm 2.5$ pN \cite{leduc.04,juli.02}, which is far greater than the
stall force of a single motor, about 6$\sim$7 pN \cite{carter.05,fisher.01}. So the extraction of membrane tube must be a result of cooperation of many motors.

Unlike the widely-studied rigid or elastic cargo\cite{juli.95,klumpp.05,wang.09}, membrane tube are actually fluid like. The lipids on the membrane tube, which the
motors are adhered to, can flow on the surface of membrane tubes. So some models suggest that only the leading motor can apply force to the membrane tube
\cite{camp.06,brug.09}. According to these models, however, it is obvious that the motors can not apply enough force to extract membrane tube from vesicles.
Camp\`{a}s \textit{et al.} have considered various motor cooperation schemes (see Ref.\cite{camp.08} for details) and suggested that there are three protofilaments
of the microtubule simultaneously involved in the pulling process. This proposal seems plausible, but no experiments support it directly. Especially, the structure
and elasticity of motors are not considered in the former models, while these details can substantially affect the collective transport of motors \cite{wang.09}. So
the cooperation mechanism should be further revised, and other mechanism could be possible.

In this paper, we focus on the extraction mechanism of membrane tube by processive motors (kinesin) and propose a two-track-dumbbell model (see Sec. \ref{S.Mo} and
Fig. \ref{fig1} for details ), taking account of the elasticity of the motor molecule and the excluded volume effect of both the head domain and the tail domain of
the motor molecule. By this model, we conclude that motors along a single microtubule protofilament can also generate enough force to extract membrane tubes from
giant vesicles.

\section{ Model} \label{S.Mo}

In the experiments of Ref. \cite{leduc.04,camp.08}, each kinesin binds to a rhodamin-labeled biotinylated lipid on the vesicle membrane through a streptavidin
molecule. And the density of kinesin on the membrane can be directly controlled by fixing the biotinylated lipid concentration in the membrane. It has been found
that if the density of kinesin on the vesicle exceeds a threshold about 100$\sim$200 $\mu$m$^{-2}$, membrane tubes can always be formed; and if the density is
smaller than 100 $\mu$m$^{-2}$, no tube extraction can be observed over a long time period. These experiments suggest that if the density is low, there are no
enough motors gathering at the tip of membrane tube, and no enough force can be generated to maintain the tube growth. So a group of kinesin on the tip is essential
for the formation of membrane tube. As mentioned in Sec. \ref{S.In}, however, it seems that only the leading one of the motors can apply force to the membrane tube,
but generate no enough force to resist the retreat of membrane. So there must exist some mechanisms, in which the non-leading motors at the tip region of membrane
tube can also apply force to the membrane tube.

\begin{figure}[b]
\includegraphics[width=8cm]{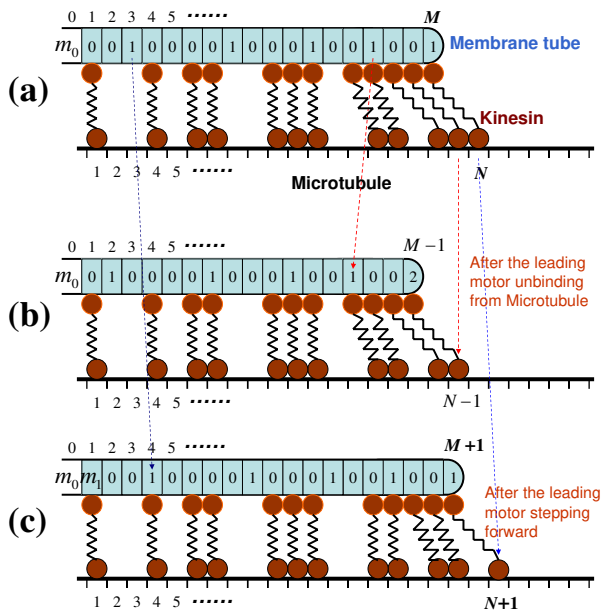}
\caption{\label{fig1} (Color online) The two-track-dumbbell model with consideration of the elasticity of motor stalk. The membrane tube is discretized as
one-dimensional lattice, with the the same lattice spacing as that of the microtubule. The number in each box represents the number of free motors in each membrane
site. (b) shows the situation after the leading motor unbinding from microtubule. Free motors are dragged back by the membrane tube and change their positions
accordingly upon the motion of membrane tube. (c) shows the situation after the leading motor stepping forward. Free motors are dragged forward by the membrane
tube. The number of free motors at site 1, $m_1$, adopts 1 with a possibility of $m_0$, else $m_1=0$, which flows from the vesicle. In (b) and (c) the actual length
of the membrane tube is determined by the equilibrium of force, while the figure just to give a demonstration.}
\end{figure}

Noticing the finite volume of the complex of motor tail domain and streptavidin. The tail domain complexes of neighboring motors may exclude each other, and
generate repulsive force in between. In this way, the non-leading motors can also apply force to the leading motor and thus to the membrane tube. On the other hand,
the elasticity of the stalk of kinesin molecule has been found to play an interesting role in multi-motor transportation in our previous study \cite{wang.09}. In
the present paper, the kinesin stalk is also considered as a spring, with an elastic constant $k_{spring}$. Combining the above two points,  one can picture the
kinesin molecule as a dumbbell with a spring-like stalk (Fig. 1). One end of the dumbbell (the tail domain) can slide on the surface of the membrane tube, and can
not detach from membrane tubes because of strong connection between kinesin and lipid. The other end of the dumbbell (the head domain) can either bind to or unbind
from the microtubule, and the bound motor can step forward with a rate $k_f$ or detach from the microtubule with a rate
\begin{equation}
 k_u(f)=k_{u0}\exp(f/f_d),
\end{equation}
where $k_{u0}$ is the unbinding rate at vanishing load, and $f_d$ is the detachment force with the value of about 3.0 pN \cite{klumpp.05}. The backward stepping of
motor is neglected here. The motors which attach to the membrane tube but not bind to the microtubule (i.e., free motors), can diffuse freely on the surface of
membrane tube with a diffusion constant $D$ and bind randomly at a rate $k_b$ to non-occupied sites on the microtubule.

To carry out a simulation of the tube growth (Sec.\ref{S.Alg}), we have adopted a discrete microscopic approach in which the membrane tube is discretized into $M+1$
sites (called membrane sites) numbered from 0 to $M$ (Fig. 1), and the lattice spacing $d$ equals to the periodicity of microtubule, 8nm. The boundary membrane site
0 is connected to the reservoir of motors (the lipid vesicle) which can supply motors continuously when membrane tube is growing, and the motor density at this site
is $\rho_0$, i.e. the motor density on the vesicle. So the number of motors on membrane site 0 is $m_0=2\pi rd \rho_0$, where the radius of membrane tube $r$ is
related to the retreating force $F$ and the bending modulus $\kappa $ of the membrane by $r=2\pi \kappa /F$ \cite{juli.02}. Since the whole perimeter of the tube
exceeds 60 nm, each  membrane site can contain many motors. In our simulations, however, we find that the number of free motors at a given membrane site seldom
exceeds two, which means the free motor density is so low that their diffusion is almost uncorrelated. So the diffusion rate of a free motor along the membrane tube
can be taken as $k_d=D/d^2$ \cite{camp.08}. A bound motor steps forward to the next un-occupied microtubule site with a rate $k_f(f)=V(f)/d$, where $V(f)$ is the
force-velocity function for single motor transport which has been widely studied both experimentally and theoretically \cite{carter.05,fisher.01}. In this study,
the $V(f)$ function is adapted from theoretical results of two-state stochastic model of Fisher and Kolomeisky, with [ATP]=1.0 mM (see Ref. \cite{fisher.01} for
details).

We now turn to the tip of membrane tube. Each bound motor behaves just like a spring connecting two beads. One bead (the head domain) binds to the microtubule,
while the other (the tail domain) connects tightly to the membrane tube. If a motor bears force, the spring is stretched. In this model, the loading force is
directly transmitted to the leading bound motor. The following motors can also share the load, by pushing the leading motor due to the exclusion of their tail
domains (Fig. \ref{fig1}). In fact, most of the tip motors would bear the load. For instance, head domains of two neighboring motors may be separated by one or more
empty microtubule sites, but their tail domains can still interact with each other since their stalks can be stretched (see Fig. \ref{fig1}(a), the five leading
motors on the tip will share the load). Motors which bear the load are called pulling motors. The length of membrane tube is determined by the equilibration between
retreating force and extracting force which depends on the distribution of bound motors on the microtubule.

If a motor unbinds from the microtubule, it can diffuse on the surface of the membrane tube (Fig. \ref{fig1}(b)). If a motor steps forward on the microtubule, its
tail domain will be dragged forward (Fig. \ref{fig1}(c)). For an unpulling motor, the tail domain is usually dragged forward by one membrane site after a forward
step, and it does not change the tube length and the state of other motors. But an unpulling motor can also become a pulling motor when it encounters another
pulling motor after a forward step. Then the load should be redistributed among pulling motors.

\section{Simulation Algorithm} \label{S.Alg}
We adopt the Gillespie algorithm \cite{gil.76} to simulate the growth process of the membrane tube, based on the model introduced in Sec. \ref{S.Mo}.

As mentioned, each bound motor can unbind from microtubule and step forward, and the free motors can diffuse and bind to microtubule. In the simulation, we define
an array $P_l$ numbered in order $l\in[1, 2N+3M+1]$, which represent all the transition rates of the system if the excluded volume effect between the motors is not
considered.
\begin{subequations} \label{eq2}
\begin{eqnarray}
P_{2i-1}=n_i k_u,&~&~~for~1\le i\le N,
\\
P_{2i}=n_i k_f,&~&~~for~1\le i\le N,
\\
P_{2N+3j-2}=m_j k_d,&~&~~for~1\le j\le M,
\\
P_{2N+3j-1}=m_j k_d,&~&~~for~1\le j\le M,
\\
P_{2N+3j}=m_j k_b,&~&~~for~1\le j\le M,
\\
P_{2N+3M+1}=m_0 k_d,&&
\end{eqnarray}
\end{subequations}
where $N$ indicates the microtubule site occupied by the leading motor, $M$ is the total length of the membrane tube. Noting that $N,M$ are not fixed numbers but
vary with the process of tube growth. $n_i$ is the number of bound motor at microtubule site $i$, which is either 0 or 1. $m_j$ is the number of free motor at
membrane site $j$. $m_0$ is the number of motor at membrane site 0, and $P_{2N+3M+1}$ represents the diffusion rate of the motor from membrane site 0 (i.e., the
reservior) to site 1.

$P_{2i-1}$ corresponds to the unbinding rate of the motor from microtubule site $i$, $P_{2i}$ corresponds to the forward stepping rate of the motor at microtubule
site $i$. $P_{2N+3j-2}$, $P_{2N+3j-1}$ correspond respectively to the backward and the forward diffusion rate of a free motor at membrane site $j$. $P_{2N+3j}$
corresponds to the binding rate of a free motor at membrane site $j$ to the microtubule.

Considering the excluded volume effects, some of the above transitions are actually prohibited, so the corresponding  transition rates $P'_l$ must be 0, or else
$P'_l=P_l$. Especially, $P'_{2N+3M-1}=0$ because the free motors at membrane site $M$ can not diffuse forward.

\begin{figure}
\includegraphics[width=8cm]{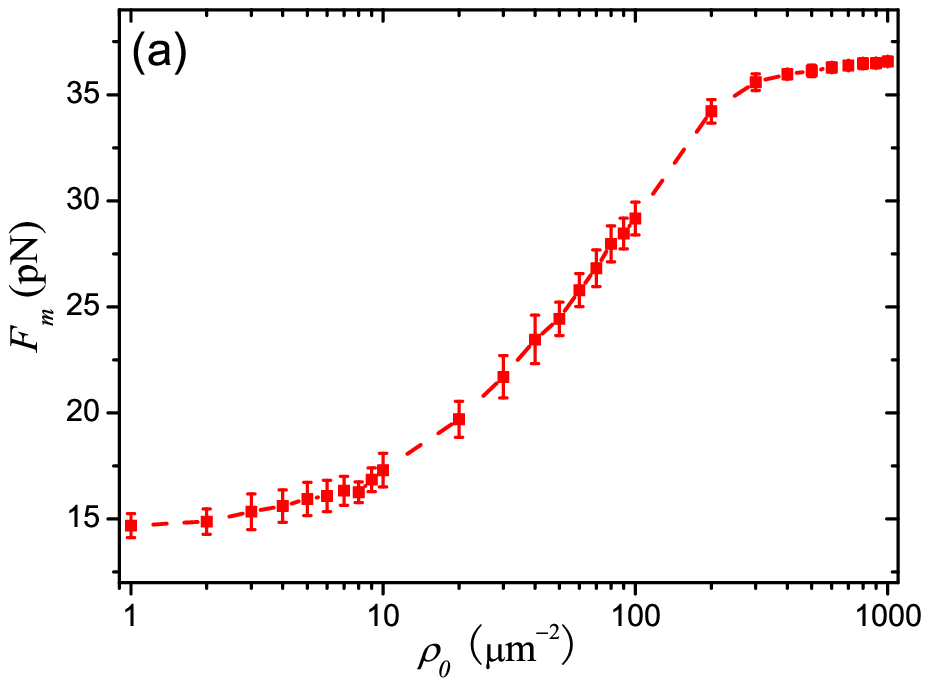}
\includegraphics[width=8cm]{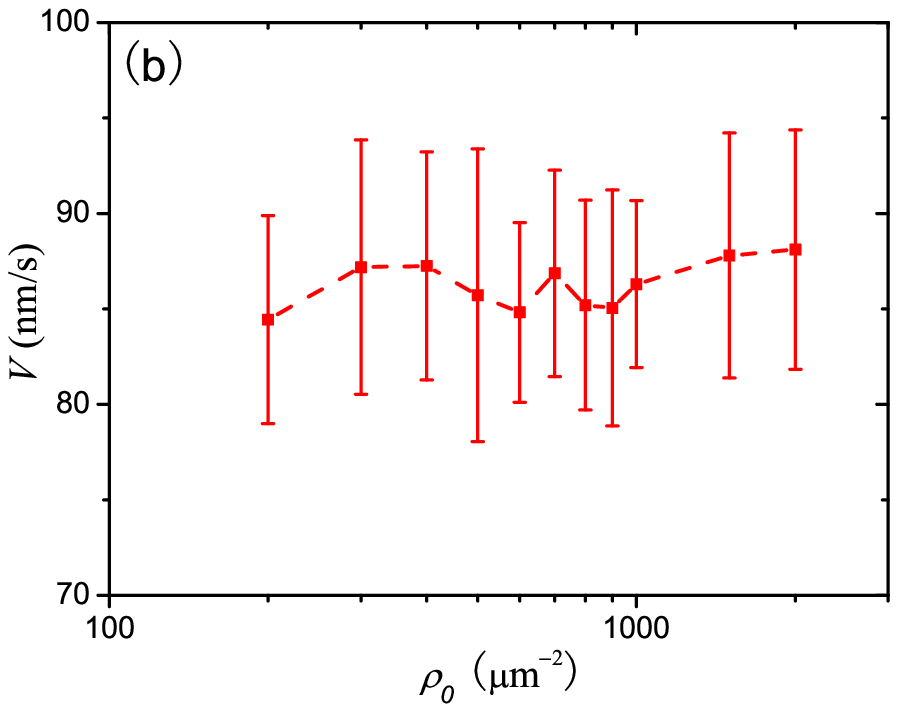}
\caption{\label{fig2} (a) The threshold force $F_m$ for membrane tube extraction as a function of the surface motor density $\rho_0$ on the vesicle. The error bars
represent standard deviation calculated from 30 repeated simulations. (b) Velocity of membrane tube growth as a function of  $\rho_0$ when the retreating force is
set as 27.5 pN. The error bars represent standard deviation calculated from 20 independent simulated membrane tubes. The simulation parameters are listed in Table
\ref{tab:table3}.}
\end{figure}

We define the global transition rate as
\begin{equation}
S = \sum_{l=1}^{2N+3M+1} P'_l.
\end{equation}
In Gillespie algorithm, the time step $\Delta t$ is not fixed and is a stochastic variable distributed exponentially with a characteristic time scale $1/S$. $\Delta
t$ determines how long to wait to see a motor (no matter on which site) performing any one of the above mentioned transitions. To determine which transition will
take place in the next time step, we generate a random number $R$ distributed uniformly in the range $[0, S]$. The largest value of $m \in [1, 2N+3M+1]$ that
fulfills the following inequality,
\begin{equation}
\sum_{l=1}^{m-1} P'_l \le R,
\end{equation}
means that the $m$-th transition numbered in Eq. \ref{eq2} will occur in the next time step.

The simulation is performed as follows. The initial length of the membrane tube is set as $M=20$ with each site occupied by a bound motor, and the actual transition
rates $P'_l$ are calculated. Then the following steps are repeated.

(1) Calculate the global transition rate $S$, and determine the stochastic time step $\Delta t$.

(2) Determine which transition to take place. In this step, one of the following four possible cases may be encountered.

a) A bound motor unbinds from the microtubule. So the number of free motors at the corresponding site of the membrane tube will increase by one. If this unbinding
motor is a pulling motor, the load will be redistributed by other motors according to the equilibration of force, and the length of the membrane tube will also be
changed too.

b) A bound motor steps forward. Its tail domain will be dragged forward. If it is a pulling motor, the load will be redistributed among pulling motors, and the
length of the membrane tube will be changed too. Sometimes an unpulling motor can become a pulling motor after a forward step, as mentioned in Sec. \ref{S.Mo}.

c) A free motor diffuses from one membrane site to another.

d) A free motor at membrane site $i$ binds to the microtubule, then $m'_i=m_i-1$, $n_i=1$.

(3) After one of the above four cases performed, the array $P'_l$ is updated with the new configuration.

(4) The time $t$ is updated to $t+\Delta t$ and the process is repeated.

The repetition stops when it reaches the conditions that we set up, which will be explained in the following.

\begin{table*}
\caption{\label{tab:table3}Parameters used in the simulations; reference sources are indicated.}
\begin{ruledtabular}
\begin{tabular}{ccccccc}
 Parameter&$k_{spring}$&$f_d$&$D$&$k_b$&$k_{u0}$
&$\kappa$\\ \hline
 Value&0.3 pN/nm&3.0 pN&1 $\mu$m$^2$s$^{-1}$&4.7 s$^{-1}$&0.42 s$^{-1}$&10$k_BT$ \\
 Ref.&\cite{cop.97}&\cite{klumpp.05}
 &\cite{leduc.04,camp.08}&\cite{leduc.04,camp.08}&\cite{camp.08,vale.96}&\cite{leduc.04,camp.08}\\
\end{tabular}
\end{ruledtabular}
\end{table*}

At a constant motor density $\rho_0$, there exist a threshold retreat force $F_m$ above which the motors cannot extract a membrane tube from the vesicle. The $F_m$
is determined as follows \cite{camp.08}. For a given value of the motor density $\rho_0$, we initially set the retreating force a very large value and repeat this
process 200 times. If the force is too large, the membrane tube may retreat completely. If membrane tube retreat can be observed in all the 200 cases, we lower the
force and repeat the process again till a threshold value $F_m$ is reached, at which at least one among the 200 membrane tubes does not retreat. Typically, membrane
tube retreat usually occurs when $M < 50$, and seldom occurs when $M >100$. In our simulation, if the length of the membrane tube exceeds $M=250$ (i.e. $2 \mu$m),
the membrane tube can grow persistently and no retreat can be observed, so we set $M=250$ as the threshold value which means the forming of a membrane tube.

To calculate the growth velocity of membrane tube, we let the membrane tube grows to a long enough length, 6.4 $\mu$m. Since the system is not in steady state at
the beginning, we calculate the average growth velocity when the length of the membrane tube is in the range from 0.4 to 6.4 $\mu$m.

\begin{figure}[b]
\includegraphics[width=8.5cm]{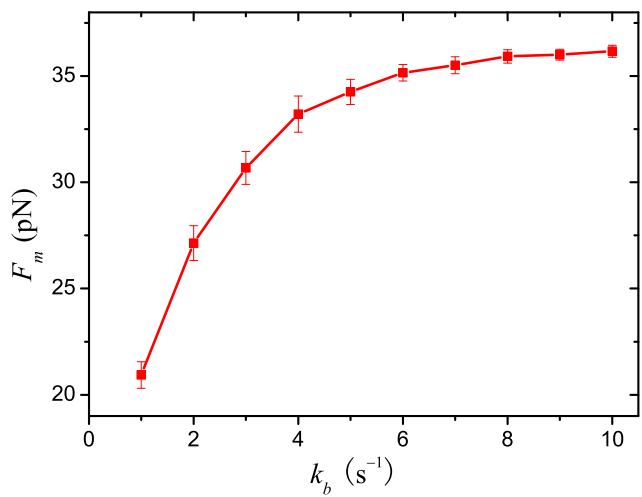}
\caption{\label{fig3} The threshold force $F_m$ for membrane tube extraction as a function of the binding rate of motor $k_b$, at the surface motor density
$\rho_0=200~ \mu $m$^{-2}$. The error bars represent standard deviation calculated from 30 repeated simulations.}
\end{figure}

\section{Results and Discussion} \label{S.Re}

\subsection{The growth of membrane tube}
In Ref. \cite{leduc.04,camp.08}, experiments showed that there exists a threshold value of motor density below which the motors cannot extract membrane tubes from
the vesicle at a given extraction force $F$. It means that at a constant motor density $\rho_0$, the motors cannot pull a membrane tube if the retreat force is
larger than a threshold force $F_m$. If $F_m$-$\rho_0$ relation is known, we can predict the threshold density above which membrane tubes can be extracted from
vesicles. In order to compare our results with Ref. \cite{camp.08}, we use the same parameters as Ref. \cite{camp.08}, which are listed in Table. \ref{tab:table3}.
And we also adopt the same method to find the threshold force $F_m$ as Ref. \cite{camp.08} has done, as mentioned in Sec. \ref{S.Alg}.

In Fig. \ref{fig2}(a), we plot the simulated $F_m$-$\rho_0$ relation. The force $F_m$ increases with $\rho_0$ from 14$pN$ (corresponding to $\rho_0=1.0~
\mu$m$^{-2}$) to 36 pN (corresponding to $\rho_0=1000~ \mu $m$^{-2}$). In the experiments, the actual extraction force for membrane tube is found about $27.5 \pm
2.5$ pN \cite{leduc.04}. From Fig. \ref{fig2}(a), we can know the corresponding threshold value of $\rho_0$ to extract membrane tubes from vesicles is about
$100~\mu$m$^{-2}$, which agrees quite well with the experiment values about 100$\sim$200 $\mu$m$^{-2}$ \cite{leduc.04,camp.08}.

In Table. \ref{tab:table3}, all the parameters but the binding rate $k_b$, are adapted from experiments. $k_b=4.7$ s$^{-1}$ is a fitting value from Ref.
\cite{leduc.04}, and its reliability would be questioned. In Fig.\ref{fig3}, we plot the $k_b$-dependence of $F_m$ at $\rho_0=200~ \mu$m$^{-2}$. $F_m$ increases
fast with $k_b$ when $k_b$ is smaller than 4.0 s$^{-1}$, and then increase slowly to a saturated value about 36 pN. When $k_b>2.0$ s$^{-1}$, $F_m$ will be larger
than 27.5 pN, which is big enough to extract membrane tubes from a vesicle.

The growth velocity of membrane tubes as a function of $\rho_0$, at the given load of 27.5 pN,  is shown in Fig. \ref{fig2}(b). The mean velocity is about $80 \sim
90$ nm/s, consistent with the experiment value $90 \pm 60$ nm/s. The almost independence of the mean velocity on $\rho_0$ is a prediction which can be tested by
further experiments.

\begin{figure}[b]
\includegraphics[width=8cm]{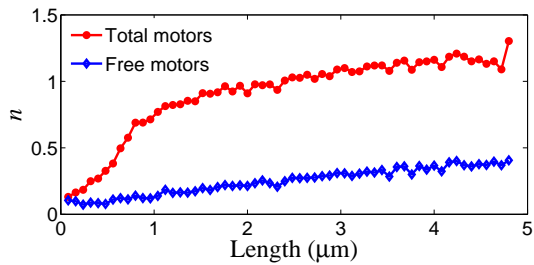}
\caption{\label{fig4} (Color online) The distribution of motors (i.e., the average motor number per site) along the membrane tube. The data are averaged from 100
independent simulated membrane tubes at $\rho_0=200~ \mu $m$^{-2}$.}
\end{figure}

\begin{figure}
\includegraphics[width=8cm]{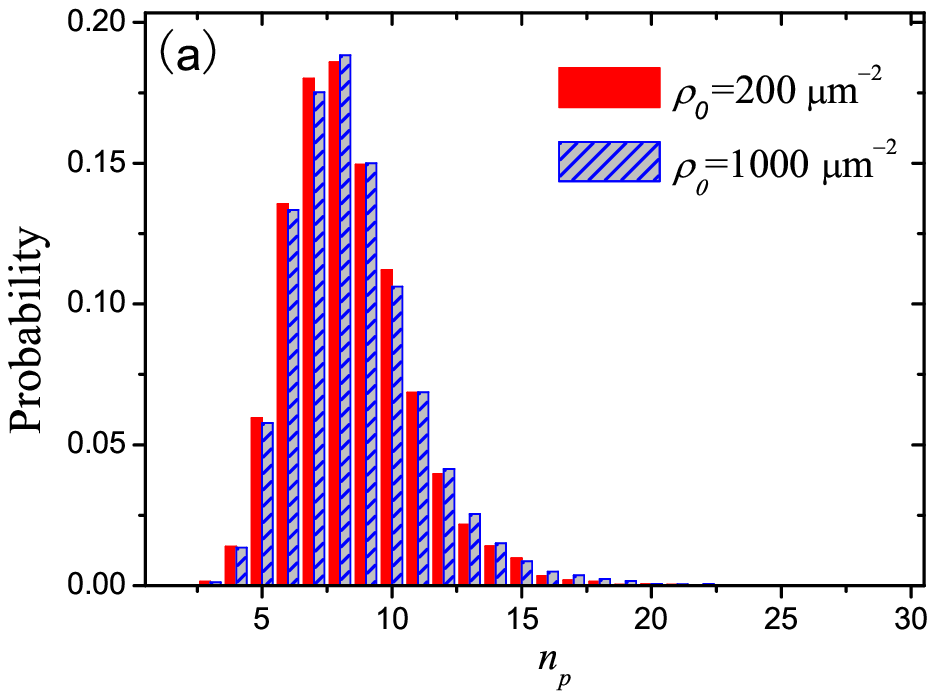}
\includegraphics[width=8cm]{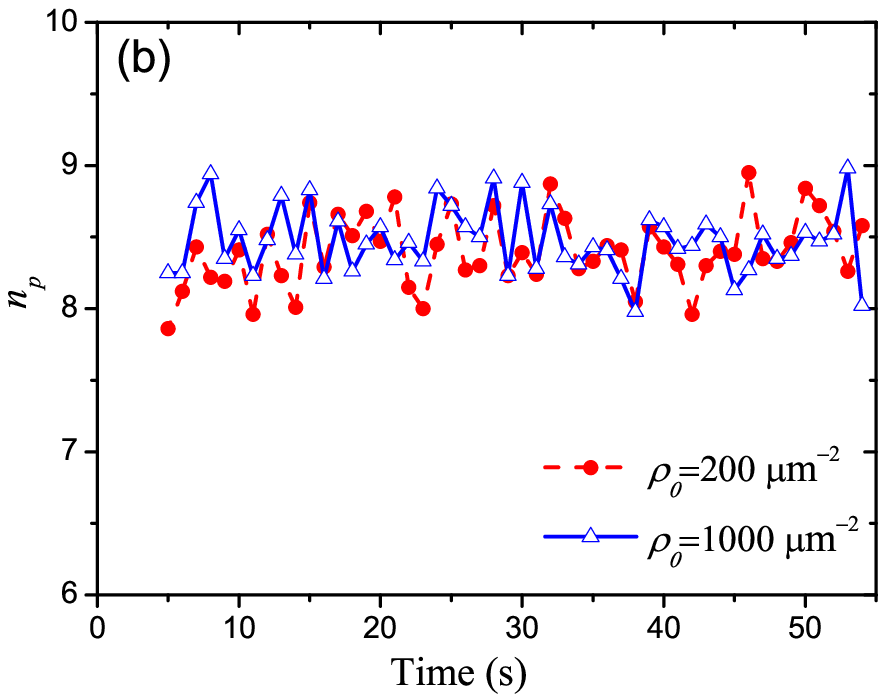}
\caption{\label{fig5} (Color online)  (a) The probability distributions of the number of pulling motors for two different motor densities, $\rho_0=200~ \mu
$m$^{-2}$ and $\rho_0=1000~ \mu $m$^{-2}$ , respectively. The data are obtained from 10000 examples. (b) Average number of pulling motors $n_p$ as a function of
time, which are averaged from 100 independent simulated membrane tubes.}
\end{figure}

\subsection{The distribution of motors}
In Ref. \cite{leduc.04}, the motors are found not uniformly distributed along the membrane tube. At the tip region, the density is bigger than that of the rest of
the membrane tube. Our simulation results are in agreement with this observation. Fig. \ref{fig4} shows that the mean density of motors increases from the vesicles
to the tip of membrane tube. Because the motors in the tip region bear the load, their velocities are smaller than the motors in the rest region, so the motors will
accumulate in the tip region and generate enough force to extract the membrane tube. Especially at the very tip of the membrane tube, almost each microtubule site
is occupied by a bound motor. Moreover, there are free motors on the membrane tube, so the motor density in this region exceeds 1 per site. Since the motors in the
tip region unbind from microtubule more frequently than that of other region due to high load, the density of free motors in the tip region is also higher than that
of other region as shown in Fig. \ref{fig4}. These free motors will diffuse and rebind to the microtubule, and then step toward the tip region again, forming a
cycle of motor flow.

At the tip region of membrane tube, the number of pulling motors is also a stochastic variable which fluctuates during membrane tube growth. We now study the number
of pulling motors and its probability distribution.

In Fig. \ref{fig5}(a) we plot the probability distributions of pulling motor number for two different motor densities, $\rho_0=200~ \mu$m$^{-2}$ and $\rho_0=1000~
\mu$m$^{-2}$ respectively. The data are obtained from 100 independent simulated membrane tubes, and 100 samples are picked out from each membrane tube at 0.5-second
intervals during its growth, so there are 10000 samples for either $\rho_0$. Since successive samples are separated sufficiently long time, they can be regarded as
independent samples, so we can plot the histogram Fig. \ref{fig5}(a). During membrane tube growth, the average number of pulling motors is 8.4 and 8.5 for
$\rho_0=200~ \mu$m$^{-2}$ and $\rho_0=1000~ \mu$m$^{-2}$ respectively. The average number of pulling motors can be regarded as a time-independent constant, as shown
in Fig. \ref{fig5}(b). So the number of puling motors is almost independent of motor density and the length of membrane tube. On average, in the growth of membrane
tube, each pulling motor bears $3\sim 4$ pN, which smaller than the stall force of kinesin, about 7pN.

\section{Summary and conclusion}
In this paper the mechanism of extracting membrane tubes by motors is studied. We propose a two-track-dumbbell model, by considering the excluded volume effect of
both the head domain and tail domain of motors, as well as the elasticity of the motor stalk. We conclude that (1) the membrane tube can be extracted by a group of
motors along a single microtubule protofilament; (2) the growth velocity of the membrane tube is almost independent of the motor densities on the vesicle $\rho_0$;
(3) the density of motors in the tip region is much higher than that in the root region.

(2) is a prediction, (3) agree with experimental data \cite{leduc.04}, and (1) is in contrary to the result of Ref. \cite{camp.08} which suggested that motors use
several protofilaments simultaneously to pull a single tube, without consideration of the structure and elasticity of the motor molecule. While these properties,
especially the elasticity of motor stalks, can affect the multi-motor transport \cite{wang.09}, it's also understandable they play a role in extracting membrane
tubes by a collect of motors. Our simulation results show that the number of pulling motors (which apply force against the retraction of membrane tube) at the tip
region of membrane tube is about $8\sim 9$ and thus they can generate enough force to extract a membrane tube from a vesicle and maintain its growth, while Ref.
\cite{camp.08} indicates the number of pulling motors is about 3 and thus three protofilaments are necessary to extract a membrane tube.

Another difference between our model and Ref. \cite{camp.08}, the choice of force-velocity (\textit{F-V}) relation of a single motor, may influence the simulation
results too. In Ref. \cite{camp.08}, the (\textit{F-V}) relation of a single motor is simplified as linear,  while we adopt the more reliable theoretical relation
from Ref. \cite{fisher.01}. In our previous study, we have shown that the property of multi-motor transport strongly depends on the \textit{F-V} relation of a
single motor \cite{wang.09}. Hence, whether the membrane tube is pulled by one row or several rows of motors, and how the intrinsic properties of kinesin motors
(the elasticity of kinesin stalk, the excluded volume effect, and the characteristic \textit{F-V} relation of a single motor) are involved in the extracting
process, should be investigated by future experiments.

\section*{ACKNOWLEDGMENTS}
Z. Q. Wang thanks the support of NSFC under Grant No. 11047188, and Chinese Universities Scientific Fund under Grant No. QN2009046. M. Li thanks the support of NSFC
under Grant No. 11075015.

\end{document}